\documentclass[amsmath,showpacs,twocolumn,aps,prl]{revtex4-1}

\usepackage{amsmath,amsfonts}
\usepackage{graphicx}
\usepackage{color}
\usepackage{enumerate}

\date{\today}

\begin{document}

\title{Localized end states  in density modulated quantum wires and rings}
\author{Suhas Gangadharaiah, Luka Trifunovic, and Daniel Loss}
\affiliation{Department of Physics, University of Basel, Klingelbergstrasse 82,
4056 Basel, Switzerland}

\begin{abstract}
We study finite quantum wires and rings in the presence of a charge density
wave gap induced by a periodic modulation of the chemical potential. We show
that the Tamm-Shockley bound states emerging at the ends of the wire are
stable against weak disorder and interactions, for discrete open chains and
for continuum systems. The low-energy physics  can be mapped onto the
Jackiw-Rebbi equations describing massive Dirac fermions and bound end
states. We treat interactions via the continuum model
and show that they increase the charge gap and further localize the end
states. In an Aharonov-Bohm ring with weak link, the bound states give rise
to an unusual $4\pi$-peridodicity in the spectrum and  persistent current as
function of an external flux. The electrons placed in the two localized
states on the opposite ends of the wire can interact via exchange
interactions and this setup can be used as a double
quantum dot hosting spin-qubits.
\end{abstract}
\date{\today}

\maketitle

\begin{figure}
  \includegraphics[width=0.85\columnwidth]{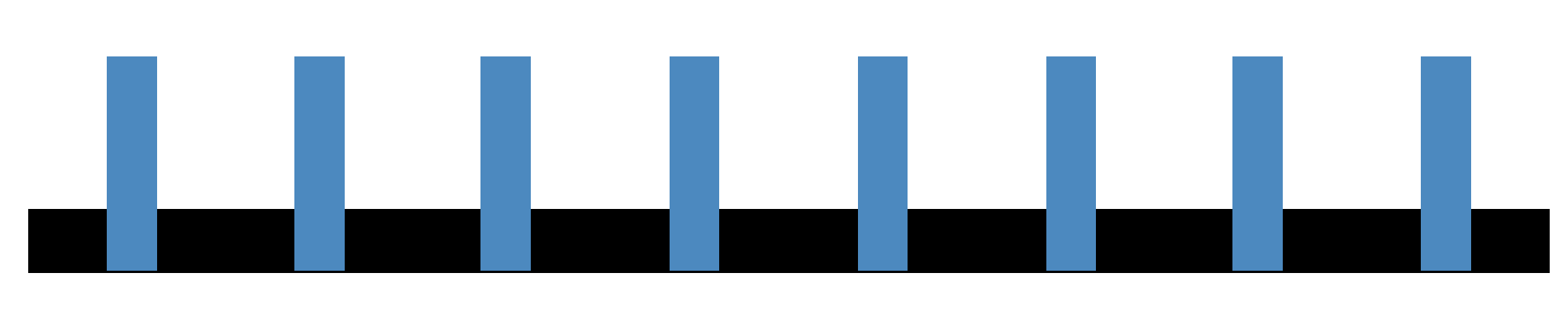}
  \caption{
The figure shows a quantum wire (black) of length $L$ with negatively charged gates (blue) forming a superlattice 
potential. Due to the induced charge density modulation a bound state at each wire end can emerge.
}
  \label{CDW}
\end{figure}

\emph{Introduction.} Over the last decades a number of proposals have been made
for solid-state implementations of a quantum computer. Among these, electron
spins in GaAs quantum dots~\cite{loss,hanson} are most promising candidates with
unusually long coherence times~\cite{bluhm2011}. Such dots contain typically
many levels which are filled according to Hund's rule. Thus, the condition
for a spin-qubit, which requires the  presence of only a single unpaired
electron, becomes challenging~\cite{hanson}, and the scalability of such an
approach is still an open problem. In this letter we propose a simple setup,
involving periodically modulated gates on top of a quantum wire (see
Fig.~\ref{CDW}), which eventually results in an effective double dot system.
Due to the spatial modulation of the gate voltage the energy spectrum of the
quantum wire acquires a charge density wave (CDW) gap. Recently, similarly
modulated setups  have been discussed with focus on  metal-insulator
transitions~\cite{malard} and transport properties in an infinite-wire
superlattice~\cite{thorgilsson}. Here, we show that the modulated quantum wire
supports localized states at each end of the wire, known as Tamm-Shockley bound
states~\cite{tamm,shockley}, with their energies lying inside the gap. These
end wave-functions are well protected from the continuum and can host stable
spin-qubits.

\begin{figure}
  \includegraphics[width=8.5cm]{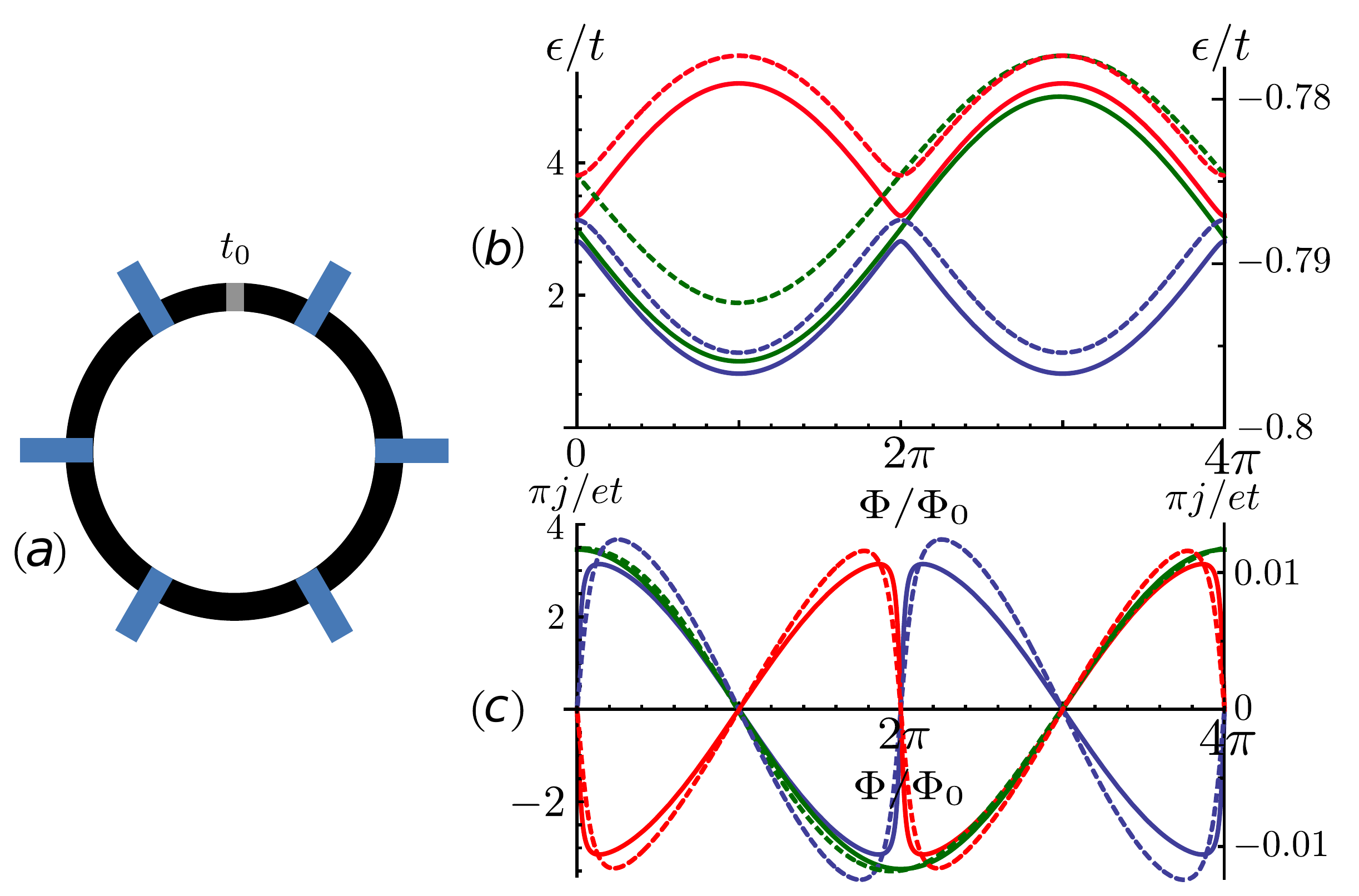}
  \caption{(a) Quantum wire (black)  in an Aharonov-Bohm-ring geometry with negatively charged gates
  (blue). The bound states are localized on  either side of the
   weak link (grey) of strength $t_0$.
    The energy (b) and persistent current $j= - \partial F/\partial \Phi$ (c) dependence on the flux $\Phi/\Phi_0$ are plotted with the solid
    curves for the effective model (Eq.~(\ref{eq:model_Hamiltonian})) and with the dashed curves for the
    lattice model (Eq.~(\ref{eq:CDW})).   The parameters
 for the red and blue  solid curves correspond to $\delta/t_0=2.2$ and $(\epsilon_{+} + \epsilon_{-})/t_0= 6.2$, and for the green curve  to  $\delta/t_0=2.0$ and $(\epsilon_{+} + \epsilon_{-})/t_0= 6.0$ (see Eq.~(\ref{energy})).
 While for the dashed curves the parameters  are   $\Delta/t_0=3.85$ (red and blue) and  $\Delta/t_0=4.54$ (green). The ratio $\Delta/t=0.5$ and $N=50$
  is the same for all three dashed curves.
 Here, $t_0$ is chosen such that we
have a degeneracy at $\Phi/\Phi_0=2 \pi$. Assuming only the lower bound
  state is filled, the persistent current shows an unusual $4\pi$-periodicity as function of $\Phi/\Phi_0$.
 }
  \label{AB_spectra}
\end{figure}

We consider one-dimensional (1D) discrete and continuum models and find a number
of remarkable features for the  end states resulting from the CDW modulation.
In particular, using exact numerical  diagonalization of the discrete open chain
we analyze the  stability of these states in the presence of a random  potential
and find that for weak disorder the end states remain stable. For the continuum
model we consider a periodically modulated potential of the form
$\Delta_0\cos(k_{CDW}x +\vartheta)$, where $\Delta_0$ is the strength of the
potential, $k_{CDW}$ the CDW vector, and $\vartheta$ a constant phase. For
$\vartheta=\pi/2$, the CDW phase supports zero energy bound states which are
remarkably robust to position dependent fluctuations in $\Delta_0$. We also show
that for $\vartheta=\pi/2$ the model  maps to the Jackiw-Rebbi model for massive
Dirac fermions with midgap bound states~\cite{jackiw}. We treat interactions via
fermionic and bosonic techniques and find that they primarily renormalize the
gap and decrease the localization length. We consider end states in a
ring-geometry  by connecting them directly via tunnel junction (see
Fig.~\ref{AB_spectra}). The  Aharanov-Bohm oscillation in such rings exhibits an
unusual $4\pi$ periodicity, providing a striking signature of the existence of
end states. Finally, we show that the two opposite end states serve as
effective double quantum dot which can be used to implement quantum computing
gates for spin-qubits.

\begin{figure}
 \includegraphics[width=1.05\linewidth]{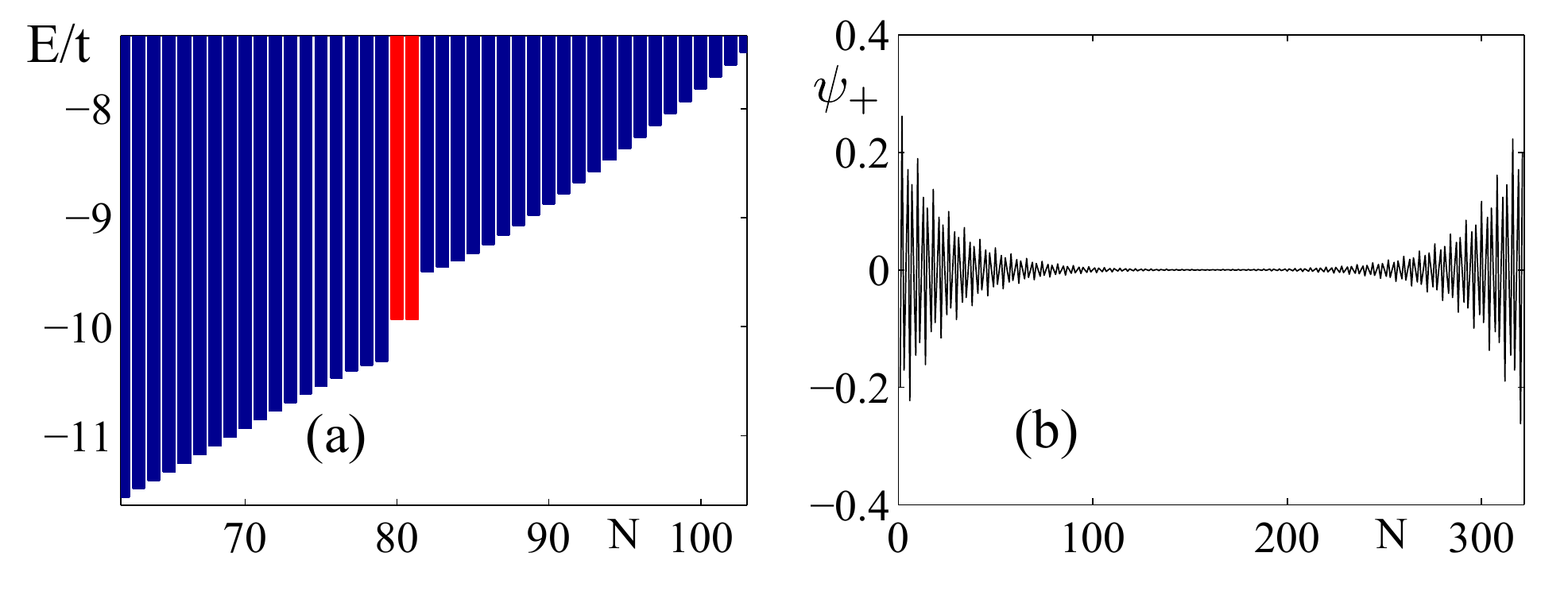}
   \caption{(a) The part of the spectra around the gap of the Hamiltonian given
   by Eq. (\ref{eq:CDW}), obtained by exact diagonalization. The red bars
   denote two almost degenerate bound (midgap) states. We have chosen for the
   parameters $t=7$, $\Delta=0.8$, and $N=320$. (b) One of two bound states.
   Plotted here is  $\psi_{+}=\psi_L+\psi_R$, where $\psi_{L,R}$ are  states
   localized at the left (right) end of the wire. }
   \label{spectra}
\end{figure}

\emph{Lattice model.}
The typical lattice  model for 1D spin-less fermions  in the presence of  CDW
modulation is described by~\cite{spinless2}
\begin{equation}
  \label{eq:CDW}
H= -t\sum_{j=1}^{N-1} \big[c^\dagger_{j+1}c_{j}+\text{h.c.}\big]  +  \Delta
\sum_{j=1}^{N} \cos\big[2 k_{CDW}  j a  + \vartheta  \big]   c^\dagger_{j}c_{j},
\end{equation}
where $c_j$ is a fermion operator at the site $j$, $N$ is the total number of
lattice sites, $t>0$ is the hopping integral, $\Delta>0$ the CDW gap,  $k_{CDW}$
the CDW wave-vector, $a$  the lattice constant and $\vartheta$ is an arbitrary
phase. The energy spectrum under the constraint of  open boundary conditions is
obtained by exact numerical diagonalization  and we find that the criterion for
the existence of bound states depends on the sign of the potential at the
beginning and end sites. For illustrative purposes we have considered $k_{CDW}=
\pi/4a$  and $\vartheta =\pi/2$, this choice  corresponds to negative potential
at the initial two sites with the overall profile  given by $\Delta
\sum_{j=1}^{N} \cos\big[j \pi/2 + \pi/2  \big]  \equiv \Delta(-1,0,1,0,-1,...)$.
If the phase of the potential is chosen such that one end of the wire has
positive whereas the other end has negative potential then only one end state is
obtained. On the other hand, for a reflection symmetric  potential profile
about the center of a long wire (with both ends having negative potential),
there will be two degenerate mid-gap states, $\psi_R$ and $\psi_L$, localized at
the right and left boundaries {\em resp.}, being the well-known Tamm-Shockley
states~\cite{tamm,shockley}. Fig.~\ref{spectra}(a) shows  the spectrum of an
$320$ site chain. Reducing  the wire length  causes  exponentially small
splitting in the energies of the bound states, with the new states described by
the symmetric and anti-symmetric combination of $\psi_R$ and $\psi_L$. We obtain
the bound states to be in the middle of the gap only when $\vartheta = \pi/2$
and $t\gg \Delta$.

\begin{figure}
 \includegraphics[width=0.95\linewidth]{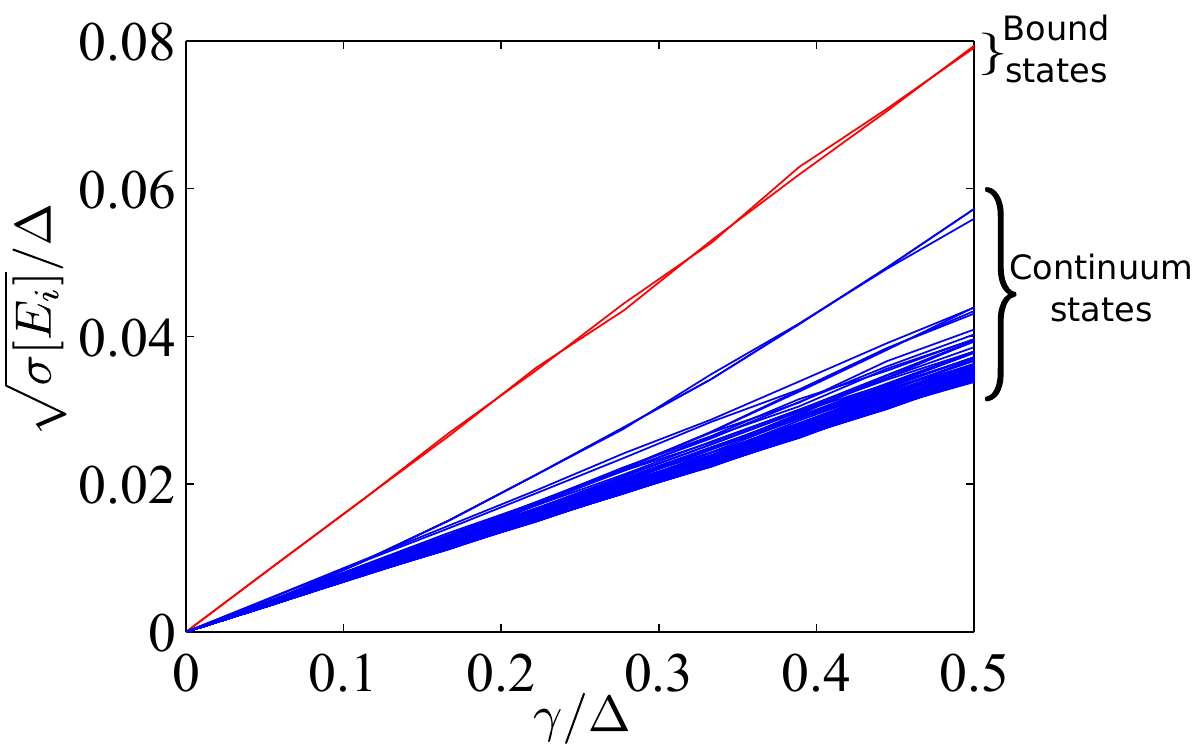}
  \caption{The dependence of the root-mean-square value $\sqrt{\sigma[E_i]}$ of
 the $i$-th energy level ($i=1\dots N$, i.e., for all energy levels) on the
 standard deviation of the random disorder potential.}
  \label{disorder}
\end{figure}

\emph{Disorder effects.} For realistic systems, some degree of random disorder
is unavoidable. To study this effect  in our lattice model, we add a random
on-site potential $\sum_i V_i c_i^\dagger c_i$.  Here, ${V_i}$ is taken
according to a Gaussian distribution with zero mean and standard deviation
$\gamma$.  Fig.~\ref{disorder} depicts the linear dependence of the
root-mean-square $\sqrt{\sigma[E_i]}$ of the $i$-th energy level ($i=1\dots N$,
i.e., for all energy levels) on the standard deviation of the random disorder
potential~\cite{note_dist}. Since the slopes of the bound states are less than
$1$, we conclude that the end states remain gapped even for disorder strengths
comparable to the gap~($\Delta$). As  $\gamma$ is increased,  Anderson
localization sets in.  We also observe as $\gamma$ is increased that the end
states begin to mix with other (spatially)  nearby localized states, thus
effectively causing the end states to be more delocalized. Additionally, it is
readily observed from Fig.~\ref{disorder} that the end states are more affected
by disorder compared to all the continuum states. The ratio thereof depends on
$\xi/L$, since this difference is coming from the spatial localization of the
end states. For a weak disorder, the aforementioned dependence is linear, while
for a strong disorder the dependence becomes more complicated due to the
emergence of Anderson localization.

So far we have considered a particular realization of the lattice model. We
next consider the  continuum case, this  limit describes the low-energy physics
of a large class of one-dimensional lattice models with CDW (or superlattice)
modulation. Recently, there has been intense activity on exotic quantum matter,
such as Majorana fermions (chargeless)~\cite{kitaev, motrunich,sau,oreg,
hassler,potter,alicea,gangadharaiah,duckheim,stoudenmire,bena} and massless Weyl
fermions~\cite{vishwanath,burkov} among others. Here, we will show that our
setup allows for the realization of the Jackiw-Rebbi Hamiltonian~\cite{jackiw},
describing a massive Dirac fermion of charge $1/2$ as end state.

\emph{Continuum model.}
We consider  a quantum wire in the presence of a gate-induced potential with
periodicity $\lambda_{CDW}=2\pi/k_{CDW}$. For carrier densities smaller than
the intraband energy gap only the lowest subband is occupied. The physics of
the fermion mode $\Psi_\sigma$  ($\sigma = \uparrow,\downarrow$ is the spin
index) in the lowest subband  is described in terms of the slowly varying right
$\mathcal{R}_\sigma(x)$ and left $\mathcal{L}_\sigma(x)$ parts and is expressed
as $\Psi_\sigma(x)= \mathcal{R}_\sigma(x)e^{ik_F x} +
\mathcal{L}_\sigma(x)e^{-ik_Fx}$. For an open wire, the boundary condition
$\Psi_\sigma(x=0)=0$ imposes the constraint~\cite{fabrizio,eggert},
$\mathcal{R}_\sigma(x)=-\mathcal{L}_\sigma(-x)$. Thus, the Hamiltonian can be
expressed in terms of right movers only.

The non-interacting Hamiltonian can be written as a sum of two parts,
$H_0=H_0^{(1)} +H_0^{(2)}$, where the kinetic part can be expressed in terms of
only the right moving fermions (the original range $[0,L]$ now becomes $[-L,L]$)
and is given by $H_0^{(1)} =-i v_F\int_{-L}^L dx
\mathcal{R}^\dagger_\sigma(x)\partial_x \mathcal{R}_\sigma(x)$ (summation on the
spin indices is assumed) and the CDW term by $H_0^{(2)} =\Delta_0 \int_0^L dx
\cos(2k_{CDW} x +\vartheta)  \Psi^{\dagger}_\sigma(x)\Psi_\sigma(x)$, with
$\vartheta$ being a constant phase factor. Thus, $H_0 = (1/2)\int_{-L}^{L}dx
\mathbf{R}^\dagger_\sigma \mathcal{H}_0\mathbf{R}_\sigma$, where
$\mathbf{R}_\sigma(x) =  [\mathcal{R}_\sigma(x), \mathcal{R}_\sigma(-x)]^T$ and
the Hamiltonian density $\mathcal{H}_0$ for each spin is the same and given by
\begin{eqnarray} \mathcal{H}_0 = -i v_F  \tau_z \partial_x  + m_1(x) \tau_x +
m_2(x) \tau_y, \label{eq:Hamiltonian-density} \end{eqnarray} where
 \begin{eqnarray}
 m_1(x) &=& -\cos[ 2\delta k x+ \vartheta \text{sgn}(x)] \Delta_0/2, {}\nonumber\\
m_2(x) &=&  \sin [2\delta k x+ \vartheta \text{sgn}(x)] \Delta_0/2,
\end{eqnarray}
and $\delta k  = k_{CDW}-k_F$.  If $\delta k =0$ and the charge-density wave
vanishes at the boundary, i.e.,  $\vartheta =\pi/2$, then it is easy to verify
that  $\mathcal{H}_0$ satisfies the `chiral symmetry'~\cite{ludwig}
$\mathcal{P}\mathcal{H}_0=-\mathcal{H}_0\mathcal{P}$ ($\mathcal{P}$ is a complex
conjugation operator). Moreover, the eigenvalue equation ($\mathcal{H}_0
\psi_\sigma=\epsilon \psi_\sigma$) of the chiral symmetric $\mathcal{H}_0$ is
related to the Jackiw-Rebbi equation describing massive
fermions,~\cite{jackiw,note}
\begin{eqnarray}
\mathcal{H}^{JR}\psi^{JR} =[\tau_z\partial_x + m \text{sgn}(x)]\psi^{JR} = \epsilon \tau_x  \psi^{JR},\label{eq:JR}
\end{eqnarray}
via the transformation, $\psi^{JR}=\mathcal{U}^{-1} \tau_y\psi$ and
$\mathcal{H}^{JR}=\mathcal{U}^{-1}\mathcal{H}_{0} \tau_y\mathcal{U}$,  where
$\mathcal{U}=\exp(i\vec{\tau}\cdot\hat{n}2\pi/3)$ and $\hat{n} =
(\hat{i}+\hat{j}+\hat{k})/\sqrt{3}$. Here, $\tau_{x,y,z}$ denote Pauli matrices
acting on the spinor $\mathbf{R}_\sigma(x)$. Solving the eigenvalue equation
for $L\gg v_F/\Delta_0$ one obtains  exponentially decaying bound states
$\psi_\sigma\sim \exp[-(\Delta_0/2v_F)x]$ and  $\psi_\sigma\sim
\exp[-(\Delta_0/2v_F)(L-x)]$ at $x=0$  and $x=L$.  Away from the chiral symmetry
point ($\vartheta \ne \pi/2$) bound states still exist as long as
$\sin\vartheta>0$, with the eigenstates given by $\psi_\sigma\sim \exp[-i
(\Delta_0 \exp[-i \vartheta]/2v_F)x]$ and $\psi_\sigma\sim \exp[-i (\Delta_0
\exp[-i \vartheta]/2v_F)(L-x)]$. For infinite wires the  eigenvalues are
degenerate  and given by  $\epsilon = -\Delta_0 \cos(\vartheta)/2$. However,
finite length introduces overlap between the end states leading to an
exponentially small splitting in the energy (see below and
Fig.~\ref{AB_spectra}).

In a realistic quantum wire the  gap $\Delta(x)$ and the phase $\vartheta(x)$
will  invariably be position dependent. Assuming this dependence to be weak, the
correction in  lowest order in  $\delta(x)/\Delta_0\ll1$ is given by,
\begin{eqnarray}
\delta \epsilon &=&-  \frac{\Delta_0 }{4v_F}\int_0^\infty dx \delta(x)\sin2\vartheta(x)
e^{-\Delta_0\sin[\vartheta_0] x/v_F},\label{eq:corrections-energy}
\end{eqnarray}
where $\langle \delta(x)\rangle=0$  and $\langle \vartheta(x)\rangle= \langle
\vartheta_0+\delta\vartheta(x)\rangle =\vartheta_0$, and they both vary slowly
on the Fermi wavelength $\lambda_F=2\pi/k_F$. Thus, $\delta\epsilon \ll
\Delta_0$, and the bound states remain stable to weak perturbations.

\emph{Interaction effects.}
In the following, we consider the effect of repulsive interactions on the end
states. For  simplicity, we consider spinless fermions with $k_F=k_{CDW}$ and
$\vartheta =\pi/2$. As usual in 1D, the interactions  can be split into forward
and back scattering parts. The former, $H_F=\pi v_F g_4\int_0^L dx
(\mbox{:$J_{R}J_{R}$:}+  \mbox{:$J_{L}J_{L}$:})$, where $J_{R} =
\mathcal{R}^\dagger(x)\mathcal{R}(x)$ and $J_{L} =
\mathcal{L}^\dagger(x)\mathcal{L}(x)$, is responsible for the velocity
renormalization~\cite{tsvelik}, $v= v_F(1+ g_4)$. On the other hand, the
backscattering part,  $H_B=\pi v_F g_2\int_0^L dx  \mbox{:$J_{L}J_{R}$:}$ at the
lowest order in interaction renormalizes the gap.  The mean-field gap
$\tilde{\Delta}(x)\propto g_2 v_F \langle
\mathcal{R}(x)\mathcal{R}^\dagger(-x)\rangle$ adds  to the externally induced
gap $m_2(x) = \text{sgn}(x)\Delta_0/2$. We note that similar to $m_2(x)$,
$\Delta(0_{+})=-\Delta(0_{-})$. This can be seen by invoking the boundary
condition, $\mathcal{R}(x)=-\mathcal{L}(-x)$, and by   expressing
$\mathcal{R}(x)=\exp(i\sqrt{4\pi}\phi_R)$ and
$\mathcal{L}(x)=\exp(-i\sqrt{4\pi}\phi_L)$ in terms of the bosonic fields
$\phi_R(x)$ and $\phi_L(x)$ which themselves satisfy~\cite{eggert},
$\sqrt{4\pi}\phi_R(0)=-\sqrt{4\pi}\phi_L(0)+ \pi $. Thus, for weak interactions
the bound states retain the same form as for the non-interacting case but with
renormalized velocity and gap. To estimate the gap size we evaluate the
self-energy, $\hat{\Sigma}$, using the unperturbed  Green's function for an
infinite wire, $G_0(i\omega,k)=(i\omega - v_Fk \tau_z -\Delta_0\tau_y/2 )^{-1}$.
In leading order, the gap renormalizes to $(\Delta_0/2)[1+ (g_2/4) \ln[
\text{min}(\Lambda,v_F L^{-1})/\Delta_0]]$, where $\Lambda$ is the band width.
Thus, the localization length,  given by $\xi= (2v_F/\Delta_0)\{1+g_4-(g_2/4)
\ln[ v_F / L\Delta_0]\}$, reduces with interaction.  In other words, due to the
repulsive interaction between the continuum and the end states, the latter
states get squeezed.

The renormalization of the gap can be more rigorously analyzed via bosonization.
Using standard procedures~\cite{giamarchi}, we obtain the following form for the
bosonic Lagrangian
\begin{eqnarray}
	&& \mathfrak{L}(x,t)= 
	\sum_{\nu=c,s}\Big[\frac{1}	{2v_{\nu} K_{\nu}}(\partial_t \phi_{\nu})^2
		-
		\frac{v_{\nu}}{2K_{\nu}}(\partial_x\phi_{\nu})^2 \Big]\\
	&&
		+
	\frac{  v_F}{2\pi a^2} \sum_{\eta=\uparrow,\downarrow}y_{\eta}\sin[\sqrt{4\pi}\phi_\eta - 2\delta k x - \vartheta ],\nonumber
\label{eq:bosH}
\end{eqnarray}
where the subscripts $c,s$ refer to   charge and spin, {\em resp.}  The
$\partial_x\phi _{c/s}$ field describes the charge/spin density fluctuations and
$\theta_{c/s}$ is the conjugated field, and $\phi_{\uparrow,\downarrow}=
(\phi_c\pm \phi_s)/  \sqrt{2}$.  The Luttinger liquid parameters  $K_{c/s}$ and
velocities $v_{c/s}$ encode interactions, and $y_{\uparrow,\downarrow}=
a\Delta_0/v_F$. The sine term denotes the coupling of up and down spin fermions
with the external potential. As before, we assume  $\delta k=0$. In general,
there are two additional terms: one of them arises due to backscattering between
opposite spin electrons and is given by  $\cos(\sqrt{8\pi}\phi_s)$,  and the
other, $\cos(4\sqrt{\pi}\phi_c -4k_F x)$, describes the Umklapp scattering.
However, both can be neglected  as the two operators flow to zero under a
renormalization group (RG) treatment.

The scaling dimensions of $\sin(\sqrt{4\pi}\phi_{\uparrow,\downarrow} )$,
$d_{\uparrow,\downarrow} =(K_c +K_s)/2\approx 1$, indicate that near
commensurability ($\delta k  v_F/\Delta_0 \ll 1$) the sine terms are strongly
relevant. The parameters $y_{\uparrow,\downarrow}$ have an identical flow [so as
to preserve the $SU(2)$ symmetry, this also implies $K_s=1$] towards the strong
coupling regime and yields an effective localization length $\xi\sim a  (a
\Delta_0/v_F)^{2/(K_c-3)}$ for the bound state. Thus as before the role of the
interactions is to reinforce the externally induced gap. We note that under RG
additional terms of the type $\partial_{i} \phi_c \partial_{i} \phi_s$ (where
$i=x,\tau$) are generated, however, they are marginal and leave  the essential
physics unaltered.

\emph{Detection.}
A viable approach for detecting the energy splitting between the bound states is
through persistent current measurements. For this the wire should be in a ring
geometry so that  the end states are connected together via a tunnel junction
and  also large enough such that the energy splitting between the bound states
is small yet the overlap of the  localized wave-functions  remain non-zero.
Such a set-up can enclose magnetic flux $\Phi$ inducing  Aharonov-Bohm  (AB)
oscillations in a mesoscopic (phase-coherent) regime.
Next consider a  single electron placed in one of the bound states.  The
effective Hamiltonian for the spinless fermion in terms of the orthogonal
symmetric, $|+\rangle$, and anti-symmetric, $|-\rangle$, states can thus be
written as~\cite{spinless2}
\begin{eqnarray}
H &=&  \sum_{\eta=\pm}\Big[\epsilon_{\eta}  + \eta t_0 \cos\Big(\frac{\Phi}{\Phi_0}\Big)\Big] |\eta\rangle\langle \eta |
+ i t_0   \sin\Big(\frac{\Phi}{\Phi_0}\Big) \nonumber\\
&&\times \Big[|-\rangle\langle +| - |+\rangle\langle -|\Big] ,\label{eq:model_Hamiltonian}
\end{eqnarray}
where $\Phi_0=h/e$ is the flux quantum, $\epsilon_{+}$ ($\epsilon_{-}$)  the
energy of the symmetric (anti-symmetric) mode, and the tunneling across the
weak link is associated with a factor $\nu t_0 \exp(i \mu \Phi/\Phi_0)$, where
$\nu,\mu=\pm 1$ and $t_0$ the tunneling amplitude. For (anti-) clockwise
tunneling we have $\mu=+$($-$), while the sign of $\nu$ depends on the relative
sign between the wave-functions across the weak link. The energy eigenvalues
are
\begin{eqnarray}\label{energy}
\epsilon_{1/2} = \frac{1}{2}\big( \epsilon_{-}
+\epsilon_{+} \pm \sqrt{4t_0^2 +\delta^2 -4 t_0\delta \cos[\Phi/\Phi_0]}\big),
\end{eqnarray}
where $\delta=|\epsilon_{+} -\epsilon_{-}| $. At $\Phi /\Phi_0 =2\pi$  the
separation between the two eigenvalues is minimal and  given by $|2
t_0-\delta|$. For large separations, the energy levels exhibit the usual $2
\pi$ dependence on the flux $\Phi/\Phi_0$. In contrast, for a flux sweep-rate
$\omega$ larger than $|2 t_0-\delta|$ a scenario emerges wherein an electron
placed in one of the levels can jump to the second level and come back to the
original one after a second $2\pi$ phase, thus exhibiting  an unusual
$4\pi$-periodicity in the persistent current, $j= - \partial F/\partial \Phi$,
where $F$ is the free energy~\cite{dloss}. By independently varying $t_0$ and
$\omega$ the splitting $\delta$ can be estimated.  For typical values
$t_0\sim\delta\sim 10\mu\text{eV}$ we estimate $j\sim 0.1 \text{nA}$, which is
of measurable size~\cite{mailly,bluhm}. For the observation, the phase-coherence
length $L_\phi$ of the ring needs to exceed $L$. For GaAs rings, we note that
$L_\phi \gtrsim \mu \text{m}$ for sub-Kelvin temperatures~\cite{mailly,bluhm}.

The effective model, Eq.~(\ref{eq:model_Hamiltonian}), does not take into
account the contribution arising from the filled Fermi sea of continuum states.
However, when the number of continuum states below the gap is even---the states
come in pairs with mutually canceling contributions to the current. On the other
hand, when this number is odd, the topmost filled continuum state contributes to
the current. Nevertheless, the amplitude of the persistent current, due to the
end and continuum states, scale differently with the lattice length $N$---the
latter behaves like $1/N$, while the former like $\delta\sim e^{-\xi/Na}$. Thus,
for chains with $N\gg1$ and $\Delta\sim \hbar v_F/Na$, the persistent current
will be dominated by the end states and our effective description fully applies.
The dashed curves in Fig.~\ref{AB_spectra} include  contributions from the bound
states as well as the filled  Fermi sea. Indeed we have confirmed  that the
contributions  from the  continuum states are two orders of magnitude less
compared to those from the bound states. Finally, for the spinfull case,  the
amplitude of $j$ simply doubles, whereas the periodicity remains unchanged.

\emph{Effective quantum dot.}
Similar to the discrete quantum dot states, the presence of  spinful,
CDW-induced, localized states in the quantum wire opens up an intriguing
possibility for the realization of a quantum computer device.  These states are
well separated from the continuum and can be filled by tuning the chemical
potential to the end state level. We note that these `quantum dots' contain
automatically only one orbital level, and no individual gates are needed to tune
them into a single electron regime. Due to incomplete screening there will be
half-filling, i.e., only one state on either end will be filled.  This is simply
because once one of the energy levels on either end is filled, to fill the
remaining two levels requires additional energy to overcome the Coulomb
repulsion. The physics of the half-filled state is described  by the usual
Hubbard model, $H=-t\sum_{\sigma=\uparrow,\downarrow}
(c^\dagger_{\sigma,R}c_{\sigma,L} + h.c.) + U
\sum_{i=L,R}n_{\uparrow,i}n_{\downarrow,i}$, where $t$ is the tunneling
amplitude and $U$ is the onsite repulsion.  For the energy hierarchy $\Delta\gg
U\gg t$ the effective Hamiltonian acquires the Heisenberg form,  $H=
J\vec{S}_R\cdot\vec{S}_L$, where $J=4t^2/U$. The effective exchange coupling $J$
can be controlled by changing the gate potential which determines the overlap
between the left and right end modes and hence the tunneling amplitude $t$. We
note that for weak  overlap, $t$ is small and $U$ large making the $J$ to be
small, whereas for strong overlap the opposite is true~\cite{kondo}. By
switching on and off the  exchange constant in an appropriate sequence, the
essential operations of the quantum dot, both the `swap'  and
`square-root-of-swap' operations can be performed, which, together with  two
single spin-qubit operations, enables the fundamental XOR gate~\cite{loss}.

Finite overlap between the right and the left end states can be ensured if their
localization  length $\xi$ is on  the order of the wire length $L$.  This
restriction yields an estimate for the strength of the periodically modulated
external voltage, $\Delta_0 \sim  \Lambda (a/L)^{(3-K_c)/2}$, where $\Lambda
\sim v_F/a$ is the band width.  A GaAs quantum wire with length $L\sim
1\mu\text{m}$ with approximately $10-20$ gates requires a Fermi wave-length
$\lambda_F \sim 50\text{nm}$. And with the   parameters~\cite{Auslaender},
$\Lambda\sim 0.2 \text{eV}$, $K_c=0.8$, and lattice spacing $a \approx 5
\text{{\AA}}$, we obtain $\Delta_0 \sim 0.04 \text{meV}$. Thus, the upper bound
for temperatures are in the achievable range of a few hundred milli-Kelvin.

\emph{Conclusion.}
We have shown that a CDW gap in a quantum wire can lead to bound states at the
ends of the wire which are stable against weak disorder and interactions. They
map to massive Dirac fermions desrcibed by the Jackiw-Rebbi model.  In an
AB-ring, the bound states lead to an unusual $4\pi$-periodicity in the
persistent current. Finally, the two opposite end states serve as effective
double quantum dot which can be  used to implement quantum computing gates for
spin-qubits.

\emph{Acknowledgements.} We acknowledge discussions with K.~Damle,
C.~Kl\"{o}ffel, D.~Rainis, B.~R\"{o}thlisberger, D.~Stepanenko, and V.~Tripathi.
This work is supported by the Swiss NSF, NCCR Nanoscience and NCCR QSIT, DARPA,
and IARPA.


\begin{thebibliography}{8}
\bibitem{spinless2}
For simplicity we omit here the spin indexes, since the $\uparrow$ and
$\downarrow$ spin channels are independent and this leads only to an
additional degeneracy.
\bibitem{note_dist}
We note that a Kolmogorov-Smirnov test shows that
the eigenenergies are not normally distributed.
\bibitem{note}
The fractional charge $1/2$ of the end states are seen only in the 
continuum model~\cite{jackiw} but not in the lattice model. This difference is 
coming from subtracting an infinite Fermi sea when passing to the continuum model.
\bibitem{kondo}
We note that because of the gap between the bound
state and the continuum,   Kondo physics does not play a role.
\end{thebibliography}
\end{document}